\documentstyle[12pt,aaspp4,flushrt]{article}
\newcommand{\axp}{1RXS~1708$-$4009}
\newcommand{\psr}{1RXS~1708$-$4009}
\def\lapp{\ifmmode\stackrel{<}{_{\sim}}\else$\stackrel{<}{_{\sim}}$\fi}
\def\gapp{\ifmmode\stackrel{>}{_{\sim}}\else$\stackrel{>}{_{\sim}}$\fi}

\begin{document}

\title{A Glitch in an Anomalous X-ray Pulsar}

\author{Victoria M. Kaspi\altaffilmark{1,2,3},
Jessica R. Lackey\altaffilmark{1}, and
Deepto Chakrabarty\altaffilmark{1}}

%% Notice that each of these authors has alternate affiliations, which
%% are identified by the \altaffilmark after each name.  Specify alternate
%% affiliation information with \altaffiltext, with one command per each
%% affiliation.

\altaffiltext{1}{Department of Physics and Center for Space Research,
Massachusetts Institute of Technology, Cambridge, MA 02139, USA}
\altaffiltext{2}{Department of Physics, Rutherford Physics Building,
McGill University, 3600 University Street, Montreal, Quebec, Canada,
H3A 2T8}
\altaffiltext{3}{Alfred P. Sloan Research Fellow}

\begin{abstract}
We report the detection of a sudden spin-up of the 11~s anomalous
X-ray pulsar 1RXS J170849.0$-$4000910 in regular timing observations
made with the {\it Rossi X-ray Timing Explorer.}  The event, which
occurred between MJD 51446 (1999 September 25) and 51472 (1999 October
21), is well characterized by an increase in the rotational frequency of
magnitude $|\Delta \nu / \nu| = (6.2 \pm 0.3) \times 10^{-7}$ and an
increase in the rate of spin down $| \Delta \dot{\nu} / \dot{\nu}| =
(1.38 \pm 0.25) \times 10^{-2}$.  These values are very similar to
those of glitches observed in the Vela radio pulsar and other young
radio pulsars. The event therefore suggests that the internal
structure of this anomalous X-ray pulsar is similar to those of the
radio pulsars.  In particular, it implies that the fractional moment of
inertia in neutron superfluid that is not corotating with the crust is
$\geq$1\%.  The detection of a glitch in this anomalous X-ray pulsar
constrains models for the origin of glitches in neutron stars.  Most
notably, it challenges models that preclude glitches in long-period
pulsars, and, under the magnetar hypothesis, suggests that large
glitches can occur in hot neutron stars.  The glitch is
consistent with the predictions of the magnetar model for anomalous
X-ray pulsars, but accretion-powered scenarios cannot be excluded
using our observations alone.
\end{abstract}


\section{Introduction}

An unusual class of X-ray pulsars, the anomalous X-ray pulsars (AXPs),
has been puzzling since the discovery of the first such object some 20
years ago (1E~2259+586, \cite{fg81}).  AXPs are characterized by spin
periods in the range of 5--12~s, steady spin down, X-ray
luminosities greatly exceeding their inferred spin-down luminosities,
steep X-ray spectra, and lack of evidence for a binary companion, either
optically or from Doppler shifts (\cite{ms95,vtv95}).  All five known
AXPs are located in the Galactic Plane, and two are coincident
with supernova remnants (\cite{fg81,gv98}).  A sixth AXP candidate is
also at the center of a supernova remnant (\cite{ggv99}).

Two main models have been suggested to explain the nature of the AXPs.
The lack of evidence for companions and their location in the Galactic
plane as well as in supernova remnants suggests that AXPs are young,
isolated neutron stars.  In this case, the steady spin-down, under the
assumption that it is due to magnetic dipole braking as in radio
pulsars, implies surface dipolar magnetic fields of
$10^{14}-10^{15}$~G.  Such fields are similar to those inferred
independently in the soft gamma repeaters; both classes of object have
therefore been suggested to be ``magnetars''
(\cite{dt92a,td95,td96a,kds+98,ksh+99}). The large X-ray luminosities
of the AXPs in this model may arise from energy from the decay of the
large magnetic field (\cite{td96a}) or from enhanced thermal emission
(\cite{hh97}).

Alternatively, it has been proposed that AXPs are accreting neutron
stars, with either (i) a very low-mass companion (\cite{ms95}) or (ii)
with no companion, but with accretion disks perhaps made of material
leftover after a companion was disrupted (\cite{vtv95}), or, for a
young neutron star, material remaining from the supernova explosion
(\cite{chn00,phn00,alp99}).  In this case, the X-ray luminosity is
from accretion, and the prolonged spin-down is a result of the pulsars
being close to their equilibrium spin period or of them being in an
extended ``propeller'' regime of centrifugal expulsion (Chatterjee et
al. 2000, \cite{alp99}).

One way to discriminate among these models is through timing
observations.  In the magnetar model, timing irregularities and sudden
spin-up events, as are seen in the young radio pulsar population, are
expected (\cite{td96a}), but long episodes of spin-up should not
be seen.  Also, a long-term periodicity superimposed on the spin-down
might be expected due to radiative precession (\cite{mel99}).  By
contrast, in an accretion scenario, large random torque fluctuations
could be expected, as might extended episodes of spin-up
(\cite{bs96,cbg+97,bcc+97}).

Past timing observations of AXPs have been hampered by poor sampling,
such that multiple interpretations of the same data set were possible
(e.g. \cite{uso93,hh99,mel99}).  Kaspi, Chakrabarty \& Steinberger
(1999) [hereafter KCS99] \nocite{kcs99} showed that with monthly
observations, phase-coherent timing of at least two AXPs (1E 2259+586,
\psr) was possible, demonstrating that the AXPs can be very steady
rotators and that such monitoring observations can in principle
distinguish among models.

Here we report on continued monitoring of the 11-s AXP
1RXS~J170849.0$-$400910 (hereafter \psr) with the {\it Rossi X-ray Timing
Explorer} ({\it RXTE}).  We show that although \psr\ rotated extremely
steadily for nearly 2~yr, a sudden spin-up event occurred between two
observations at epochs MJD 51446 and 51472.  We show that the
properties of the event are very similar to the ``glitches'' seen in
young radio pulsars.

\section{Observations and Results}
\label{sec:obs}

The {\it RXTE} observations described here are a continuation of those
reported by KCS99.  We refer the reader to that paper for details of
the analysis procedure.  Briefly, all observations were obtained with
the Proportional Counter Array (\cite{jsss+96}), with events in the
range 2.5--5.4~keV selected to maximize signal-to-noise ratio.  Data
have been obtained roughly monthly since 1998 January and were reduced
using software designed to handle raw spacecraft telemetry packet
data.  They were binned at 62.5~ms resolution and reduced to the solar
system barycenter in barycentric dynamical time using the JPL DE200
solar system ephemeris.

The spacing of the observations was carefully chosen to permit
absolute pulse phase determination using standard radio pulsar
techniques.  The timing ephemeris of KCS99 was the starting point in
the continuing analysis, with individual observations folded at the
predicted barycentric period.  A total of 64 pulse phase bins were
used.  Folded profiles were cross-correlated in the Fourier domain
with a high signal-to-noise ratio average profile in order to
determine an average pulse arrival time.  Resulting arrival times were
then analyzed using the TEMPO pulsar timing software
package.\footnote{http://pulsar.princeton.edu/tempo}

\begin{figure}
\plotfiddle{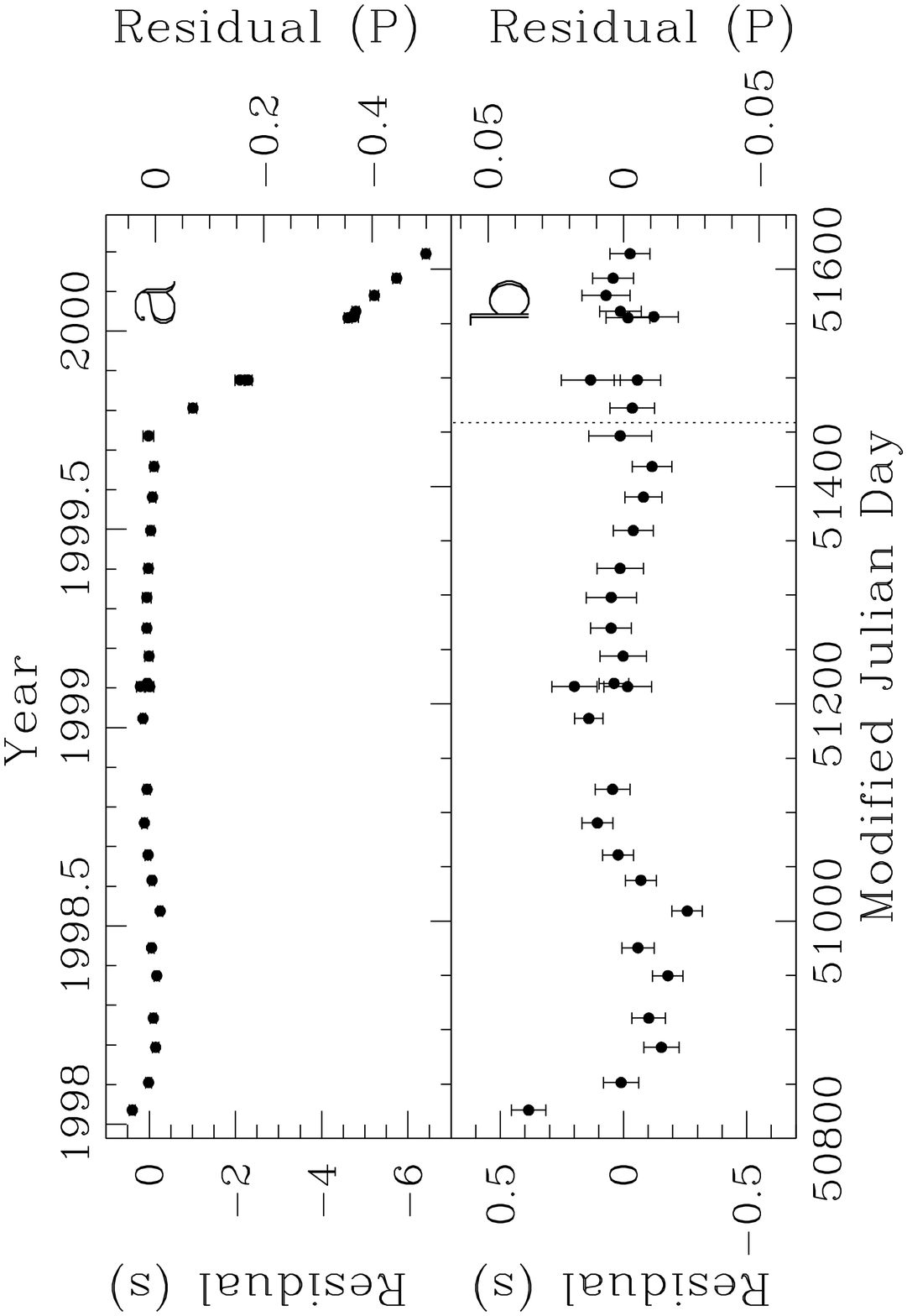}{5in}{270}{70}{70}{-280}{450}
\figcaption[fig1.eps]{Timing residuals for \psr.  (a) Residuals after
subtraction of the pre-glitch model from all pulse arrival times.  The
spin-up event, evident as the series of early pulse arrival times
relative to the long-term spin down, occurred between MJDs 51446 and
51472. (b) Residuals after subtraction of the pre-glitch model from
the pre-glitch data, and the post-glitch model from the post-glitch
data (see Table~1).  The dotted line indicates the pre- and
post-glitch separation.  Note the difference in vertical scales in (a)
and (b). \label{fig:res}}
\end{figure}

The ephemeris given by KCS99, which was determined from 19
observations made in the interval MJD 50826 -- 51324 (1998 January 13
-- 1999 May 26), continued to predict phase for over 120 days, until
MJD 51446 (1999 September 5).  This is clear from the pre-glitch
timing residuals (see Figure~\ref{fig:res}) which have RMS 130~ms
(0.012$P$, where $P=1/\nu$ is the pulse period).  The subsequent
observation, on MJD 51472 (1999 October 21), was not well-predicted,
and the following residuals grew steadily (see Figure~\ref{fig:res}a).
For this reason, we initiated a pre-planned series of three closely
spaced observations in order to independently determine the new pulse
frequency $\nu$.  All observations from MJD 51472 onward are well
modeled by a single $\nu$ and $\dot{\nu}$.  This revised ephemeris has
now properly described 9 arrival times obtained over 142~days, with
RMS residuals of only 71~ms (0.006$P$).  Table~\ref{ta:parms}
summarizes the spin parameters before and after the event, where the
values are extrapolated to MJD 51459, the midpoint between MJDs 51446
and 51472.  Residuals after subtraction of the pre-glitch model from
the pre-glitch data and the post-glitch model from the post-glitch
data are shown in Figure~\ref{fig:res}b.

The frequencies given in Table~\ref{ta:parms} imply that the pulsar
suddenly spun up, with fractional frequency change $|\Delta\nu/\nu| =
(6.2 \pm 0.3) \times 10^{-7}$.  Furthermore, following the event, the
spin down rate increased in magnitude by $|\Delta \dot{\nu}/\dot{\nu}|
= (1.38 \pm 0.25)\times 10^{-2}$.  In both cases, the uncertainties
are derived by combining those of the pre- and post-glitch values in
quadrature.  These changes are very similar to those observed in the
Vela radio pulsar, as well as in other radio pulsars of ``adolescent''
age (e.g. \cite{ml90,kmj+92,sl96,lkb+96,wmp+00}; see
\S\ref{sec:disc}).

Although the timing event is well described by a simple step function
model, it can in principle also be described by a continuous model
with a single $\nu, \dot{\nu}$ and significant $\ddot{\nu}$.  However,
in this case, the timing residuals show strong systematic trends,
including a clear discontinuity at the epoch of the event, and the RMS
residual is approximately three times larger than that in the
pre-glitch model.  Smooth deviations from a simple spin-down law have
been observed in many, if not most radio pulsars and are known as
``timing noise'' for lack of a better term.  However, discrete events
like the one we have observed for \axp, especially since they are
always observed to be sudden spin-ups, are a distinct phenomenon
classified as glitches (see \cite{lyn96} for a review).  The
identification of discrete events as a distinct phenomenon in radio
pulsars has grown out of many years of phase-coherent timing
observations of hundreds of sources, something unavailable for AXPs.
Thus, by the conventional operational definition for glitches in radio
pulsars, and by Occam's Razor, we conclude that the timing event we
have observed in \axp\ is indeed a glitch.  However, it should be kept
in mind that it may instead represent a new phenomenon not seen in
radio pulsars.  Only continued timing observations will settle this
point with certainty.

%For each pulse profile, in addition to determining a pulse arrival
%time for use in the timing analysis, we also search for pulse profile
%variations.  This is done by comparing each profile with the average
%profile by first phase aligning using cross-correlation, then
%normalizing the profiles to have equal areas, and finally fitting out
%a DC offset and scale factor.  The aligned and scaled profile is then
%subtracted from the average profile, resulting in profile residuals
%for each observation.  A $\chi_{\nu}^2$ figure of merit is then
%determined by taking the sum of the squares of the residuals, and
%dividing by $\nu$, the number of degrees of freedom (in our case $64 -
%4 = 60$).  Higher values of $\chi_{\nu}^2$ indicate profiles that
%deviate from the average profile more.  Figure~\ref{fig:chi} shows the
%values of $\chi_{\nu}^2$ versus time.  Note that taken as a whole, a
%Kolmogorov-Smirnov test indicates that the measured $\chi_{\nu}^2$'s
%are consistent (59\% probability) with having come from a
%$\chi_{\nu}^2$ distribution.  Thus, we find no strong evidence for
%pulse profile variations.  However, three of the four
%highest values of $\chi_{\nu}^2$ occurred in the data immediately
%following the spin-up event.  Considering only those profiles, the
%probability of them having been drawn from the same $\chi_{\nu}^2$
%distribution is only 2.8\%.  Thus there is perhaps some suggestion
%for pulse profile variations having accompanied the timing event,
%although we cannot be certain.

We detected no change in the 2.5--5.4~keV X-ray flux from the pulsar
at the time of the glitch.  We set an upper limit on flux variations
of $<$20\% (3$\sigma$) of the mean flux.  We also detected no
statistically significant change in the X-ray pulse profile at the
time of the glitch.

\section{Discussion}
\label{sec:disc}

The spin-up event we have observed in \psr\ is very similar to the
glitches seen in the Vela radio pulsar and other radio pulsars of
comparable age, that is, with $10^4 < \tau_c < 10^5$~yr, where
characteristic age $\tau_c \equiv P/2\dot{P}$
(e.g. \cite{sl96,wmp+00}).  In such young pulsars, observed glitches
are dominated by frequency steps of size $\Delta\nu/\nu \simeq 10^{-7}
- 10^{-6}$.  Furthermore, such glitches frequently show increases in
the magnitude of the spin-down rate of order a few percent, sometimes,
but not always, with subsequent relaxation back to the pre-glitch
value on time scales of several hundred days.  The rates of occurrence
of such glitches vary from source to source, with some occurring more
frequently than once per year (e.g. PSR J1341$-$6220,
\cite{kmj+92,wmp+00}), and most (generally the older pulsars, $\tau_c
\gapp 50$~kyr), never having been observed to glitch.  All these
properties are consistent with those of the spin-up event in \psr\
($\tau_c = 9$~kyr), namely the magnitude of the glitch, the change in
the slow-down rate, and even, very crudely, the rate of occurrence,
once per $\sim 2$~yr of observation.  Some glitching radio pulsars,
especially the well-studied Vela pulsar, have also shown significant
relaxation on time scales of hours to days (e.g. \cite{cmnp93}).
However, such behavior is on too short a time scale to be detectable
in our observations of \psr.

Large glitches in radio pulsars have been ascribed to sudden unpinning
of superfluid neutron vortices (\cite{ai75,acp89,accp93}).  The
neutron star spins down under the influence of an external torque
which acts on the crust.  For radio pulsars, the torque is magnetic
dipole braking.  Neutron superfluid in the stellar interior, which is
not well coupled to the crust, has its angular momentum carried in
quantized vortices.  The superfluid can spin down by outward motion of
these vortices.  However, vortex line pinning to crustal nuclei can
impede their outward motion.  The crust and superfluid components
therefore develop a differential angular velocity.  Occasionally,
sudden unpinning of vortex lines can occur, and the previously
decoupled superfluid can spin down, transferring angular momentum to
the crust in the process.  A spin-up event is therefore observed.  The
neutron superfluid thus acts as an angular momentum reservoir to fuel
glitches. The similarities in the properties of the spin-up event seen
in \psr\ to those seen in the Vela-like pulsars suggests that a
similar mechanism is at work in \axp.

In contrast, smaller glitches observed in the younger Crab
pulsar are dominated by changes in spin-down rate rather than in
pulse frequency (\cite{lps93}).  These are ascribed to
changes in the neutron star ellipticity due to cracking of the crust.
The magnitude and frequency of the Vela-like glitches are incompatible
with such a model but agree well with the vortex-line unpinning
model, in which the fractional angular momentum change per glitch is
roughly constant from source to source (\cite{ab94}).

From the observed glitch parameters, we can estimate the fraction of
the neutron star moment of inertia in neutron superfluid that is not
corotating with the crust, $I_s$.  First, one can show (e.g. \cite{bel99a})
that
\begin{equation}
\frac{I_s}{I_c} \geq \frac{\overline{\nu}}{|\dot{\nu}|} A,
\end{equation}
where 
$I_c$ is that of the crust and all other coupled components,
$\overline{\nu}$ is
the average spin frequency over the observing span, and
$A$ is the activity parameter (\cite{ml90}), where
\begin{equation}
A = \frac{1}{t} \sum_{i} \frac{\Delta\nu_i}{\nu}.
\end{equation}
Here, $t$ is the observing span, and the sum is over all observed
glitches.  As we have observed only one glitch for \axp, we can only
crudely estimate $A$, under the assumption that we were not extremely
lucky in detecting the glitch, and perhaps also that the small value of
$\ddot{\nu}$ observed before the glitch (KCS99) was due to relaxation
following a glitch that occurred before our observations began
(cf. \cite{lkb+96}).  Hence, we take $t \sim 3$~yr, so $A \sim 5 \times
10^{-7}$~yr$^{-1} ( 3 \; {\rm yr}/t)$, and $I_s/I_c \geq 0.01 ( 3 \;
{\rm yr}/t)$, similar to that found for many Vela-like pulsars (see
\cite{bel99a} and references therein).  Alpar et al. (1993)
\nocite{accp93} suggested a different estimate for $I_s$, namely
$I_s/I_c \geq \Delta\dot{\nu} / \dot{\nu}$, where short-term transient
contributions to $\Delta\dot{\nu}$ have been omitted.  Since we were
not sensitive to short time scale transients, our measured
$\Delta\dot{\nu}$ can be used directly, and yields $I_s/I_c \geq
0.01$, consistent with the first estimate.

Thus, the glitch implies that $I_s$ in \axp\ is similar to that in the
Vela-like pulsars.  We note, as pointed out to us by I. Wasserman
(personal communication) that this renders models of long time-scale
precession in AXPs (\cite{mel99}) unlikely, because of the expected 
dynamics of the superfluid interior (\cite{sha77,ao87}).

Ruderman, Zhu \& Cheng (1998) \nocite{rzc98} have suggested that the
origin of the vortex line unpinning events is cracking of the neutron
star crust under stresses imposed by outward-moving magnetic flux
tubes.  These tubes move because they interact with the
outward-migrating angular momentum vortex lines as the neutron star
spins down.  However, this model predicts that glitch activity should
be absent in neutron stars having $P \gapp 0.7$~s because the vortex
motion is too slow to cause the necessary stresses.  This is in
contradiction with the large glitch in the 11-s \axp.  Thus our
observations suggests that the Ruderman et al. (1998) model is
inapplicable to the glitch in \psr.  Given the similarity of this
event to those seen in Vela-like pulsars, this may cast doubt on the
relevance of the model to those sources as well.

Usov (1993) and Heyl \& Hernquist (1997) argued that spin-down
irregularities in other AXPs (1E~2259+586 and 1E~1048.1$-$5937)
are also due to glitches.  The data they used did not involve phase
coherent observations as did ours, and so their conclusions are much
less certain.  Furthermore, the fractional amplitude of the glitches
they inferred are several orders of magnitude larger than what we have
observed for \axp. Given the glitch in \axp, one might suspect that
the previous claims of glitches in other AXPs were correct.  However
our ongoing observations of AXPs 1E~2259+586 and 1E~1048.1$-$5937 do
not support the conclusion that the timing irregularities in those
objects are due to sudden spin-up events.  A detailed discussion of
these sources will be presented elsewhere.

Glitches in AXPs were predicted in the magnetar model
(\cite{td96a}). These authors argued that crust fracture and superfluid
vortex line unpinning play a major role in outbursts of soft-gamma
repeaters (SGRs), and are ultimately due to the stresses imposed on
the crust by the large magnetic field.  
%Indeed surface temperature
%variations from energy release near the fracture sites are also
%predicted in this model; these have clearly been detected as
%significant pulse profile variations in the X-ray pulsations from SGR
%counterparts following outbursts (e.g. \cite{ksh+99}).
However, the glitch alone does not provide proof of the magnetar
hypothesis.  The origin of neutron-star glitches in the vortex-line
unpinning models is independent of the source of the external torque
acting on the crust.  Rather, it relies upon an angular velocity
differential between the crust and that portion of the superfluid that is
effectively decoupled from the crust.

It has been argued (\cite{rud76,anp85,rud91c}) that the different
nature of the glitches in the very young Crab pulsar ($\tau_c =
1$~kyr, Lyne et al. 1993) and the absence of glitches in the young PSR
B1509$-$58 ($\tau_c=1.6$~kyr, \cite{kms+94}) imply that giant
glitches do not occur in the youngest pulsars because they have higher
internal temperatures, which allow a more plastic flow of vortex
lines.  However, in the magnetar model, the X-rays are a result of
thermal processes, either magnetic field decay (\cite{td96a}) or
enhanced thermal emission from initial cooling (\cite{hh97}).  In
either case, the neutron star is very hot.  This is supported by the
X-ray spectrum of \axp: it can be fit with power-law and blackbody
components (although the latter is not strictly required), which
suggest a surface temperature of $kT \simeq 0.4$~keV (\cite{snt+97}).
This is hotter than is observed in any of the Vela-like pulsars, and
higher than expected in the very youngest pulsars for all cooling
models (\cite{oge95a}), even if the measured temperature is an
overestimate of the true surface temperature because of atmospheric
effects (e.g. \cite{mpm94}).  Thus, in the magnetar model, the glitch
in \axp\ argues that the differences in the glitching behavior of the
youngest radio pulsars compared to the ``adolescent'' Vela-like
pulsars may not be primarily due to the difference in internal
temperature.

In the recently proposed AXP model in which these sources are
accreting from disks of material formed after the supernova explosion
(Chatterjee et al. 2000, Perna et al. 2000, Alpar 1999), the spin-down
rates are a result of accretion torque, which presumably acts only on
magnetic field lines anchored in the crust.  Thus, glitches might be
expected in this model as well.  In this case, the frequent glitches
seen in Vela-like pulsars might be less influenced by their age than
by their relatively large spin-down rates.  A prediction of this
hypothesis, independent of AXP phenomenology, is that glitches occur
in neutron star X-ray binaries, although large fluctuations in
accretion torque (e.g.  \cite{bcc+97}) make them difficult to detect.
The low-mass X-ray binary 4U~1626$-$67, in which the spin-down is
extremely stable apart from episodes of sudden torque reversal
(\cite{cbg+97}), should be an excellent candidate for the detection of
glitches, although none has been seen in $\sim$5~yr of timing using
the BATSE instrument.

Recent deep infrared observations of the field containing the AXP
1E~2259+586 (\cite{hkvk00}) have not detected any emission from a
putative accretion disk, casting some doubt on the fallback disk
model.  If \axp\ is indeed isolated, however, the glitch is consistent
with the magnetar model, which provides the required external torque
via magnetic dipole braking.  However, since constraining
optical/infrared observations of the \axp\ field have yet to be done,
an accretion scenario for this source cannot be ruled out.

Finally, glitches will complicate the determination of braking indexes
in AXPs as well as the search for periodic variations in the spin
period predicted in the magnetar model due to precession.
Nevertheless, continued long-term monitoring of \axp\ is essential for
the determination of the amplitude distribution and frequency of its
glitches.  Similar observations of other AXPs are necessary to
determine whether glitch behavior is ubiquitous.

\bigskip
We thank Evan Smith and the {\it RXTE} operations team for their
skill and support in scheduling the AXP monitoring program.  We also
thank A. Alpar, R. Duncan, A. Lyne, D. Nice, S. Thorsett and
I. Wasserman for useful discussions, and F. Crawford and D. Fox for
helpful comments on the manuscript.  This work was supported in part
by a NASA LTSA grant (NAG5-8063) to VMK.

%\bibliographystyle{/nfs/janeway/h1/vicky/doc/tex/apj1c}

%\bibliography{/nfs/janeway/h1/vicky/doc/tex/psrrefs/journals1,/nfs/janeway/h1/vicky/doc/tex/psrrefs/modrefs,/nfs/janeway/h1/vicky/doc/tex/psrrefs/psrrefs,/nfs/janeway/h1/vicky/doc/tex/psrrefs/crossrefs}

\clearpage

\clearpage
\begin{deluxetable}{l|cc}
\tablecaption{Measured Spin Parameters for \psr. \label{ta:parms}}
\tablewidth{0pt}
\tablehead{
\colhead{Parameter} & \colhead{Pre-Glitch Value} & \colhead{Post-Glitch Value} }
\startdata
Spin Frequency, $\nu$ (Hz) & 0.0909136408(7)  &  0.090913697(3)  \\
Spin Frequency Derivative, $\dot{\nu}$ ($10^{-13}$ s$^{-2}$) & $-1.5681(2)$
& $-1.590(4)$ \\
Spin Period, $P$ (s) & 10.99944949(8) & 10.9994427(3)  \\
Spin Period Derivative, $\dot{P}$ ($10^{-11}$) & 1.8972(3) & 1.9241(48) \\
Epoch (MJD) & 51459.0  & 51459.0 \\
RMS Timing Residual (ms) & 130 & 71 \\
Number of Arrival Times & 23  & 9  \\
Start Observing Epoch (MJD) & 50826  & 51472 \\
End Observing Epoch (MJD) & 51446 & 51614 \\
\enddata

\tablecomments{Numbers in parentheses represent 1$\sigma$ uncertainties
in the last digit quoted.}

\end{deluxetable}

\end{document}